\documentclass[aps,twocolumn,showpacs,preprintnumbers,amsmath,amssymb,floatfix]{revtex4}
\usepackage{graphicx}
\usepackage{epsfig}
\begin{document}
\def\be{\begin{equation}}
\def\ee{\end{equation}}
\def\bearr{\begin{eqnarray}}
\def\eearr{\end{eqnarray}}
\def\tc{$T_c~$}
\def\bis2{$\rm BiS_2$~}
\def\spx{$\rm 6p_x$~}
\def\spy{$\rm 6p_y$~}
\def\laob{$\rm LaOBiS_2$~}
\def\laofh{$\rm LaO_{0,5}F_{0.5} BiS_2$~}
\def\laofx{$\rm LaO_{1-x}F_{x} BiS_2$~}
\def\half{$\rm \frac{1}{2}$~}

\title{RVB States in doped Band Insulators from Coloumb forces:\\
Theory and a case study of Superconductivity in BiS$_2$ Layers}

\author{ G. Baskaran}

\affiliation
{The Institute of Mathematical Sciences, C.I.T. Campus, Chennai 600 113, India \&\\
Perimeter Institute for Theoretical Physics, Waterloo, ON, Canada}

\begin{abstract}
Doped band insulators, HfNCl, WO$_3$, diamond, Bi$_2$Se$_3$, \bis2 families, STO/LAO interface, gate doped SrTiO$_3$ and MoS$_2$ etc. are unusual superconductors. With an aim to build a general theory for superconductivity in doped band insulators we focuss on \bis2 family, discovered by Mizuguchi et al. in 2012. While maximum Tc is only $\sim$ 11 K in \laofx, a number of experimental results are puzzling and anomalous; they resemble high Tc and unconventional superconductors. Using a two orbital model of Usui, Suzuki and Kuroki we show that the uniform low density free fermi sea in \laofh is \textit{unstable towards formation of next nearest neighbor Bi-S-Bi diagonal valence bond} (charge -2e Cooper pair) and their \textit{Wigner crystallization}. Instability to this novel state of matter is caused by unscreened nearest neighbor coulomb repulsions (V $\sim$ 1 eV) and a hopping pattern with sulfur mediated diagonal next nearest neighbor Bi-S-Bi hopping t' $\sim$ 0.88 eV, larger than nearest neighbor Bi-Bi  hopping, t $\sim$ 0.16 eV.  Wigner crystal of Cooper pairs quantum melt for doping around x = 0.5 and stabilize certain resonating valence bond states and superconductivity. We study few variational RVB states and suggest that \bis2 family members are latent high Tc superconductors, but challenged by competing orders and fragile nature of manybody states sustained by unscreened Coulomb forces. One of our superconducting state has d$_{xy}$ symmetry and a gap. We also predict 2d \textit{Bose metal or vortex liquid} normal state, as charge -2e valence bonds survive in the normal state.
\end{abstract}

\maketitle

\section{Introduction}

After the discovery of high Tc superconductivity in cuprate family \cite{BednorzMuller} in 1986, efforts to synthesize new superconductors have continued and is gaining momentum. There are notable successes in these efforts, even though Tc has not surpassed the limit $\sim$ 134 K, set by a trilayer cuprate at ambient pressure. It is interesting that in most of the newly discovered low and high Tc superconductors there are strong signals for electron-electron coulomb interaction based mechanisms for superconductivity.  In this background experimental observations of superconductivity in doped band insulators HfNCl, WO$_3$, diamond, Bi$_2$Se$_3$ families, STO/LAO interface, gate doped SrTiO$_3$, MoS$_2$ \cite{HfNCl,WO3,Diamond,Bi2Se3,STOLAO,gateSTO,gateMoS2} etc. have come as a surprise. The \bis2 family of band insulators with low density of doped careers, discovered in 2012 by Mizuguchi and collaborators \cite{Mizuguchi2012,Awana2012,MizuguchiReview2014,BrianMappleReview,Kuroki2012} belongs to this category. 
Even at optimal Tc $\approx$ 11 K, only a low density of carriers $\sim 10^{19} /cm^3$ get added to the two empty conduction bands. Electron-phonon interactions should suffice to explain the observed low Tc. However, various experiments seem to point that something different is going on. 

With an ultimate aim to provide a general theory of unusual superconductivity in a class of doped band 
insulators \cite{HfNCl,WO3,Diamond,Bi2Se3,STOLAO,gateSTO,gateMoS2}, we focuss on a representative \bis2 family in the present article. Key result of the present article is that unscreened short range coulomb interactions among doped carriers in this low career density system and orbital degeneracy destabilize free fermi sea and form charge -2e valence bonds organized as generalized Wigner crystals. We explicitly demonstrate this instability for an interacting model Hamiltonian for \laofh, built on a 2 orbital model \cite{Kuroki2012} of Usui, Suzuki and Kuroki (USK model). Further we show formation of certain resonating valence bond (RVB) states, from quantum melting of charge -2e Wigner crystal.

Wigner crystals, originally discussed for fermi sea in a jellium background \cite{WignerCrystal}, contain ordered lattice of spatially localized electrons. This state is possible in low carrier density system, when coulomb forces start to dominate, as a result of reduced metallic screening. Our \textit{generalized Wigner crystal} is defined for tight binding band insulators containing relatively low density of doped carriers. Depending on the system they could form i) ordered lattice of charge -2e valence bonds or ii) ordered array of doped Mott-Hubbard electron chains or iii) stacked 2d electron sheets. Further it could be electrically insulating or conducting. We illustrate our proposal and theory using \bis2 layer family as an example. Extensive experimental studies that already exist in \bis2 family gives us interesting phenomenological clues to make theoretical progress. 

In what follows, we first discuss structure of \laob, quantum chemistry and summarise the USK model \cite{Kuroki2012} for the BiS$_2$ layers.  Then we discuss in some detail experimental results that seem to be at odds with a simple rigid band filling band picture for various properties. 

Using insights from these experimental results and physics of unscreened coulomb interaction for low carrier density systems, we build on the USK model, using \laofh, a case of optimal and \textit{commensurate doping}, as a reference system. We show that a large unscreened nearest neighbor coulomb repulsion V $\sim$ 1 eV and a favorable hopping pattern in the conduction band help destabilize the uniform fermi sea. Hybridization of Bi and S orbitals in the BiS plane is such that the sulfur mediated diagonal next nearest neighbor Bi-S-Bi hopping t' $ \sim $ 0.9 ev is larger than the nearest neighbor Bi-Bi hopping t $\sim$ 0.16 eV. Thus doped electrons in \laofh i) avoid nearest neighbor occupancy, hence coulomb repulsion V and ii) at the same time form a strong next nearest neighbor, diagonal Bi-S-Bi valence bond (figure 3,4). Through suitable ordered occupancy of 6p$_x$ and 6p$_y$ type Wannier orbitals and quantum fluctuations, the system forms generalized Wigner crystals of valence bonds and prepares ground for unsuspected resonating valence bond (RVB) states and superconductivity. 

We simplify the model Hamiltonian of USK and present in some detail three such lower energy states in the neighborhood of doping x = 0.5 : I) Wigner crystal of charge -2e Cooper pair (Bi-S-Bi diagonal valence bonds), II) quantum melted Wigner crystal of plaquette bond resonances (PBR) and III) parallel ordered array of weakly coupled doped Mott-Hubbard chains.

We compute energy of variational states I and III exactly for our model Hamiltnian, using known Hubbard dimer solution and 1d Lieb-Wu solution. We estimate energy of the state II, using perturbative arguments.

In state II delocalization is two dimensional and along the Bi planes. This will be an isotropic 2d Kosterlitz Thouless superconductor within the ab-plane. We also show that valence bond resonance within a plaquette favor an order parameter with d$_{xy}$ symmetry. It is gapful as the nodal lines pass between disconnected fermi surfaces.

In state III electrons delocalize along X or Y chains and form weakly coupled doped Mott Hubbard chain array, using the large Bi-S-Bi hopping matrix element. For a range of doping around x = 0.5 we have doped Mott insulating chain containing RVB supercondutivity correlations \cite{PWAScience87,BZA,BZAIran}. Anisotropic 2d uperconductivity gets stabilized by interchain pair tunneling \cite{WHA,Kivelson,StrongClarkeAnderson}. In this picture superconductivity will be an anisotropic 2d Kosterlitz-Thouless superconductor within the ab-plane.

At the moment we are unable to distinguish the 2d isotropic and anisotropic superconducting states within the accuracy of our microscopic theory. We suspect that both types of superconductivity occur in diffirent family members and at different parts of the phase diagram, or they could even coexist. There is also some experimental evidence for this, as we will see.

Our phases, sustained by unscreened short range coulomb interactions, are fragile. There are underlying spontaneous breaking of rotational and translational symmetry in the square lattice. Superconductivity and other orders owe their existence to an organized orbital order. In the Hubbard chain scenario, state III, normal state is a highly anisotropic metal even within the ab-plane, depending on the orientation of the Mott chain arrays.  Correpsondingly local superconductivity is also anisotropic within the 2d plane. Doped 1d Mott insulating chains, belonging to a class of Luttinger liquids, have several intrinsic instabilities: CDW, SDW and superconductivity. For a range of interaction parameters, even p-wave superconductivity is possible. It will be interesting to see if such a possibility is present for some range of physical parameters; available experimental results \cite{muSR1,muSR2,SWavePatnaik,SWave2016} have been interpretted as providing evidence for gapped s-wave superconducting state. In scenario II, normal state is likely to be a 2d Cooper pair metal or Bose metal \cite{BoseMetal}, a vortex liquid \cite{vortexMetalPWAOng}.

Several theoretical works \cite{BCSPhonon1,BCSPhonon2,BCSPhonon3,BCSPhonon4,BCSPhonon5} provide electron-phonon interaction based 
BCS mechanism for superconductivity. Other works \cite{Elbio,ChineseTheory1,ChineseTheory2,ChineseTheory3,ChineseTheory4}, including 
the original work of USK \cite{Kuroki2012}, emphasize importance of coulomb interaction based spin fluctuation mediated superconductivity 
and possibility of uncoventional superconducting order parameters. However, they work with rigid band filling (an uniform state with no orbital
order/polarization) as a starting point and study superconducting instabilities arising from coulomb interaction. \textit{Ours, on the 
other hand, at the very start reorganizes a rigid band filled fermi sea into generalized Wigner crystals} using orbital degeneracy, and 
avoids a strong nearest neighbor repulsion. We have a basic building block, namely the Bi-S-Bi diagonal spin singlet or valence bond. 
Our correlated many body state contain an optimal gain in delocalization and correlation energy. Effects of spin-orbit 
coupling \cite{SpinOrbit}, known to be important in the heavy Bi atom based systems, will be discussed in a later publication.

The present paper, which is based on phenomenology, microscopic, quantum chemical and physical arguments, is an attempt to  show new possibilities in doped band insulators. We suggest novel mechanism for superconductivity and certain unusual orders. Our estimates of superconducting Tc's are very crude. However, they point towards possibility of higher Tc's.
The message is that with suitable materials engineering in the \bis2 family, to avoid competing charge ordered phases, one could reach higher superconducting Tc's. Our mehcanism of a strong charge -2e pairing also suggests anomalous normal state describable as Bose metal or vortex liquid \cite{BoseMetal,vortexMetalPWAOng}.

\section{Crystal and Electronic Structure}

\laob is a representative member of the parent compound of the undoped band insulating \bis2 family. LaO layers and \bis2 layers are alternately stacked along c-axis \cite{Mizuguchi2012}. Electron hopping between layers along c-axis is weak.  LaO layer consists of square lattice oxygen layer sandwitched by two La square lattice layers, such that we get a 2d network of edge sharing La$_4$O octahedra. This insulating charge reservoir layer has a nominal valency La$^{3+}$O$^{2-}$.

\bis2 layer consists of two layers of edge sharing square pyramids made of 5 sulfur atoms. At the base center of each pyramid is a Bi atoms. Bases of edge shared pyramids form a BiS square lattice with Bi and S atoms forming two square sublattices. Sulfur atom at the apex of the pyramid is bonded directly to the Bi atom at the base center. Bases of the pyramids of the two layers face each other and form two ab-plane staggered BiS square planes. The two BiS planes are weakely coupled electronically \cite{KurokiBilayer}. 

Nominal valency of \bis2 layer is Bi$^{3+}$(S$^{2-})_2$. The S$^{2-}$ ion has a filled shell. Bi$^{3-}$ ion has an electronic configuration [Xe]4f$^{14}$5d$^{10}$6s$^2$. The two lowest degenerate \spx and \spy empty orbitals of Bi, after suitable hybridization with orbitals of the bridging sulfur atom, form the two lowest empty conduction bands in the BiS square lattices of the \bis2 layer. These two degenerate Wannier orbitals, with \spx and \spy symmetry, form the basis for the 2 orbital tight binding model of USK \cite{Kuroki2012}. Because of strong hybridization through the S atom at the middle of elementary plaquettes of Bi atoms, the degenerate Wannier orbitals orient and hybridize maximally along two diagonal direction of the square lattice (figure 1). Two oriented orbitals with positive and negative slopes will be referred to as X and Y orbitals respectively (figure 2). 

\begin{figure}
\includegraphics[width=0.3\textwidth]{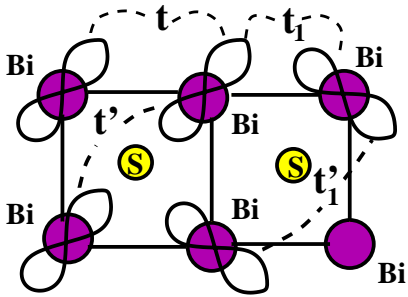}
\caption{Four types of hopping processes involving the \spx and \spy orbitals of Bi atoms, denoted by t',t, t'$_1$ and t$_1$, in decreasing strength. A sulfur atoms bridges two Bi atoms and provides the largest hopping matrix element t' $\approx$ 0.88 eV along Bi-S-Bi doagonal.} \label{fig1}
\end{figure}

In the tight binding parametrization of USK (figure 1), 4 important matrix  elements are: t' = 0.88 eV, t'$_{1}$ = 0.094, t = -0.167 and t$_{1}$ = 0.107. Matrix elements t' and t'$_1$ are between two next nearest neighbor Bi atoms bridged by a S atom; t' is between Wannier orbitals directed along the diagonal direction Bi-S-Bi; t'${_1}$ is between Wannier orbitals directed perpendicular to the diagonal direction Bi-S-Bi. Matrix elements t and t${_1}$ are between two  nearest neighbor Bi atoms with orbitals parallel or mutually perpendicular to each other.

As discussed in detail in references \cite{Kuroki2012}, if we keep only the largest diagonal hopping term, t' $\approx$ 0.88 and ignore the rest, rigid band filling scenario becomes simple. Free part of the Hamiltonian in this case describes a collection of disconnected tight binding electronic chains running along the two diagonal directions \cite{Kuroki2012,chainBrazil}. We call them X and Y chains (figure 2). In the X-chain electrons hop between next nearest neighbor \spx orbitals with a matrix element t'. Similary for the y-chains. Enery dispersion relation for X and Y chains are
\bearr
\epsilon_X(k_x,k_y) = 2t' \cos(k_x + k_y)\\
\epsilon_Y(k_x,k_y) = 2t' \cos(k_x - k_y)
\eearr
Band filling in this limit is simple and featureless. However, incorporation of t, t$_1$ and t'$_1$, the interchain hopping parameters, modify dispersion relations, resulting in a well discussed topological change in fermi surface shape as one goes through a van Hove singularity as x is increased from 0.0 to 0.5. It should be pointed out that topological phase transition generated by the weak interchain hopping occur only at small fermi energies, where a significant modification of the fermi sea, from the strong unscreened coulomb interactions, is expected.

Two of the largest coulomb interaction parameters, which have been considered by other authors as well \cite{Kuroki2012,Elbio,ChineseTheory1,ChineseTheory2,ChineseTheory3,ChineseTheory4}, will play important role in our theory. They are the onsite Hubbard U and nearest neighbor repulsion V. We approximate the inter and intra orbital two electron repulsion at a given site by a single U $\sim$ 2 eV. Similary we assume a single V $\sim$ 1 eV, for two electron repulsion between nearest neighbor sites, independent of the orbitals they occupy. 

Another important off-diagonal two body Coulomb term represented by J$_{\rm p} \approx \frac{V}{4} \sim $ 0.25 eV, rotates a diagonal spin singlet pair (figure 5) in an elementary Bi plaquette, by an angle $\frac{\pi}{2}$. An inter orbital Hund coupling at a given Bi site can be ignored, as proabability of such occupancies are very low in our low carrier density systems.

Larger value of U and V has been suggested in the literature. We find that even with our conservative values of U and V, rich physics emerge. 

Having gotten an idea about the electronic structure we will move on to a summary of the puzzling phenomenology. It guides us to build a model and suggest physics beyond rigid band filling.

\section{Puzzling Experimental Results}

In this section we briefly summarise experimental results and point out certain puzzling aspects. We begin with ARPES \cite{arpesHHWen1,arpesHDingSmallFS,arpesOrbOrder,arpesFSTopology,arpesJapanese}. For low doping, one sees small Fermi pockets, as predicted by theory. However the pockets do not close and resemble Fermi arcs seen in underdoped high Tc superconductors. Further some experiments find that volume of the fermi pockets do not agree with doping density x, indicating a possible failure of Luttinger theorem.  Quasi particle width, as given by ARPES are very broad, with widths comparable to underdoped cuprates.

A recent ARPES study \cite{arpesOrbOrder} brings out orbital poarization content in different parts of the BZ. This has been interpreted to mean that underlying Bloch states of the fermi sea have different orbital content, due to a strong 1d character of the band. We suggest that this data is not inconsisent with real space domains containing ordered doped Mott-Hubbard electronic chains (domains of generalized Wigner crystal). This interpretation has some support from STM results that we will see shortly.

ARPES also sees \cite{arpesFSTopology}, as doping is changed, an indication for change in fermi surface topology, as in a rigid band filling scenario. However, spectral functions are very broad. We interpret the seen topology change as an interesting, but residual feature of a strongly affected reference fermi sea, rather than signal for a fermi liquid. Similar situation arises in cuprates, where even in the Mott insulator and lightly doped Mott insulator one sees residual features of a strongly affected non interacting fermi surface.  Fermi liquid, as we know in metals for example, has a sharp feature with quasi particle width vanishing as T$^2$ at low temperatures. Electron-phonon interaction, important in these low carrier density material, could explain the broad spectral features through polaronic effect. However it is difficult to explain other features, as in the case of cuprates.

In a soft Xray photo emission study \cite{SoftXRay} a strong supporession of the fermi edge is seen in a \bis2 family member. The authors bring out an interesting similirity of this suppression to that seen in strongly correlated systems such as cuprates.

STM studies \cite{stmStripe,stmLargeGapHHWen,stmGapSpreadHHWen} for different family members exist. A recent observation \cite{stmStripe} of strong local orbital ordering and checker board pattern of such local orders is very interesting. Caution has been made that this could be a surface effect, rather than a bulk phenomenon. However, such a behavior in the bulk, is consistent with our generalized Wigner crystallization supported array of doped Mott-Hubbard chains and/or ordered plaquette valence bond resonances. 

Giant superconducting fluctuations, resembling pseudo gap of cuprates have been also reported. 
In one of the STM studies \cite{stmLargeGapHHWen}, what is believed to be a superconducting gap does not close well above Tc, reminescent of underdoped cuprates. A large variation in the superconducting gaps seen \cite{stmGapSpreadHHWen}in one experiment is also reminescent of underdoped cuprates. It is also noteworthy that a large value $\frac{2\Delta}{k_BTc} \approx 16$ seen in some experiments \cite{stmLargeGapHHWen} is significantly different from weak coupling BCS expectations.

Tunneling asymetry has been argued to be an important consequence of an underlying Mott character in doped Mott insulating system \cite{TunnAsymOngAnderson}. The work reported in reference \cite{stmStripe}, clearly shows a marked tunneling asymetry. We interpret this as signal for presence of self organized doped Mott insulator character.

Neutron scattering results \cite{neutron1,neutron2} show presence of strong local structural (pyramid) distortions. The results remind one of large octahedral rotations and local distortions in CuO$_2$ planes seen in single layer cuprate material such as LSCO or 2201 BISCO, where the superconducting Tc is considerbly low. These local distortion or rotations help charge localization and favor competing orders such as charge stripes, spin stripes etc., at the expense of superconducting order \cite{competingOrder,GBCompetingOrder}.

In one of the \bis2 family members, a signal for CDW transition \cite{CDW250K} around 250 K is reported. It will be interesting to see if this is connected to our proposal of generalized Wigner crystallization that supports superconducting RVB states at low temperatures.

Controlled study \cite{Doping1,Doping2} has been performed in Sr$_{1-x}$La$_x$FBiS$_2$ as a function of doping x. There is a good indication of a first order insulator to superconductor, as a function of x.  In our opinion this points out that charges in the insulating phase (below the critical x) are organized in special ways (Cooper pair Wigner crystals or self trapped valence bonds from a strong electron-phonon interactions) that enable a first order transition to a superconducting state.  Detailed STM and ARPES study in these systems are required.

Hydrostatic pressure and uniaxial pressure experiments \cite{pressureMaple,pressureAwanaReview,pressureArumugam,pressureHHWen,uniaxialPressure} also show first order phase transition, between two superconducting states having different Tc's. We suggest that they represent transition between two different self organized RVB phases supported by different patterns of generalized Wigner crystallization.

Optical conductivity \cite{Optics} in one of the \bis2 systems exhibit an upper Hubbard band like maxima centered around
1.5 eV. This interesting feature, according to our interpretation to be discussed later, provides an evidence for presence of strong Bi-S-Bi valence bonds in the ground state.

With this background in mind we will discuss our model and show that coulomb interaction can organize conduction electrons in special ways and give rise to an electron correlation based superconductivity.

\section{Instability of the Uniform Fermi Sea Towards Cooper Pair Wigner Crystal Formation}

In this section we start with the 2 band model of USK and show that even a conservative value of coulomb interaction parameters encourage orbital order and  modifies physics in an interesting way. The commensurate doping x = 0.5, in \laofh offers us a convenient place to illustrate importance of coulomb interactions and our generalization Wigner crystallization of charge -2e valence bonds. First we estimate the total energy per site in the rigid band fllled scenario. Then we construct few RVB states with lower energies. 

We consider the many chain Hamiltonian H$_c$ for a single Bi square lattice. It contains two parts: X and Y chain hopping terms and onsite and nearest and neighboring coulomb interactions U and V. To begin with we ignore interchain hopping terms t'$_1$, t and t$_1$, as they  are small compared to the largest t'. Their important effects will be studied latter. The many chain model hamiltonian is:
\bearr
H_c &=& -t'\sum_{\langle ij;X;nnn \rangle} c^\dagger_{i X\sigma} c^{}_{j X\sigma} -t'\sum_{\langle ij;Y;nnn \rangle} c^\dagger_{iY\sigma} c^{}_{jY\sigma} + \nonumber \\
&+&\frac{U}{2} \sum_i n_i^2  + V \sum_{\langle ij; nn \rangle} n_i n_j ,
\eearr
where $n_i \equiv \sum_{\sigma} (n_{iX\sigma} + n_{iY\sigma})$ is the total number of electrons at site i. The summation `nn'  `nnn' stands for nearest neighbor and next nearest neighbor (diagonal) hopping respectively.
We will construct few variational wave functions and compute their energy exactly for the many chain model Hamiltonian H$_c$ (equation 3). This will help us to show an instability of the uniform free fermi sea. At the filling x = 0.5 under consideration, mean number of electron added per site is \half. So every X and Y chain is \textit{$\frac{1}{8}$-th filled} for each spin species (figure 2). This uniform state has full orbital and lattice symmetry.

\begin{figure}
\includegraphics[width=0.3\textwidth]{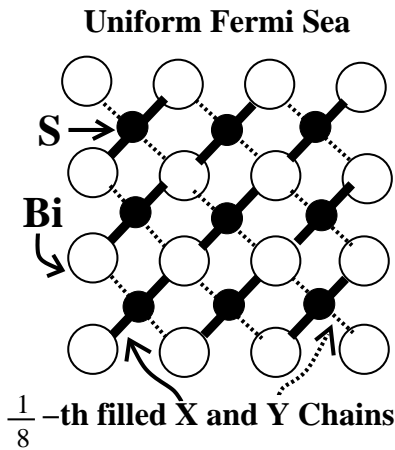}
\caption{All chains of type X and Y have $\frac{1}{8}$ filled bands for commensurate filling x = 0.5 in \laofh. This uniform state, obtained in a rigid band filling picture, is a shallow fermi sea}\label{fig2}
\end{figure}

Band width of the 1d chains is 4t'. Kinetic energy per site, counting contributions from two types of chains and two spin species is:
\be 
KE_0 = 2 \times 2 \times \frac{2t'}{2\pi} \int^\frac{\pi}{8}_{-\frac{\pi}{8}} \cos k dk \approx - t'
\ee
Hartree energy of this uncorrelated state can be estimated in the presence of coulomb repulsions U and V.  A given Bi site, containing two orbitals can have a total 16 many electron states = 1 (empty) + 4 (one electron + 6 (two electron) + 4 (three electron) + 1 (completely filled 4 electron) states.  Probability of finding a 2 electrons at a given site for the filling x = 0.5 is  $ (\frac{1}{8})^2 \times 6$. Here the first term is the probability of finding two electrons in a single site and the second term is the number of two electron states.  We ignore three and four electron states, which occur with lower propabilities, $ (\frac{1}{8})^3$ and $ (\frac{1}{8})^4$. So energy cost from onsite repulsion is $ \approx \frac{6}{64} U$. Similarly energy cost (Hartree energy) per site from nearest neighbor coulomb repulsion from four neighbors is $ (\frac{1}{8})^2 \times 4^2 \times\frac{4}{2} \times $ V $ = \frac{V}{2}$. In the decoupled chain limit under consideration, Hartree-Fock contribution from V term is zero.

Thus kinetic plus interaction energy per Bi site for the fermi sea, with complete orbital symmetry, for x = 0.5 is 
\be
E_{0.5} \approx  -t' +  (\frac{6}{64} U + \frac{1}{2} V) \approx - ~ 0.2 ~ eV
\ee
Inclusion of interchain and other hopping matrix elements do not produce appreciable correction to the variational energy. We observe that major energy increase in this low density system comes from the nearest neighbor repulsion V. 

\begin{figure}
\includegraphics[width=0.25\textwidth]{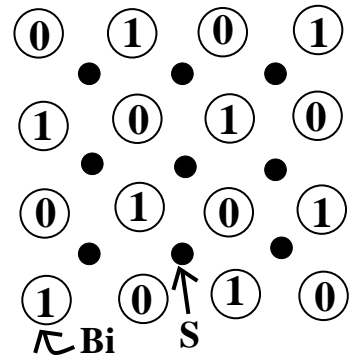}
\caption{A \textit{classical Wigner crystal} ground state for \laofh with filling x = 0.5. One of the sublattices of the Bi lattice, marked by occupancy `1' contains one added dopant electron at each site. Degeneracy of this classical ground state is 4$^\frac{N}{2}$, arising from two fold orbital and two fold spin degeneracy at every occupied site. N is the number of Bi atoms in the BiS plane.} \label{fig3}
\end{figure}

Let us consider a classical electron crystal state, which avoids coulomb repulsion U and V completely (for x = 0.5), through single electron occupancy of one of the sublattices of Bi square lattice and zero occupancy in the other (figure 3). This crystalline state of electrons is highly degenerate. The 4$^\frac{N}{2}$ fold degeneracy arises from the two fold orbital and two fold spin degeneracy. 

We will remove degeneracy of the above classical ground state step by step, to see the emergence and importance of the diagonal Bi-S-Bi valence bond. In the first step we form charge -2e Cooper pairs or valence bonds as shown in figure 4. It is a Wigner lattice of ordered and frozen diagonal valence bonds, where no two electrons are nearest neighbors.

We calculate energy of a Bi-S-Bi diagonal valence bond (charge -2e state) using the ground state energy of a two site Hubbard model (Hubbard dimer) 
\be
E_{\rm vb} = - \frac{1}{2}[U^2 - 16t'^2]^{\frac{1}{2}} + \frac{U}{2}.
\ee
This expression is very instructive. Since ratio of band width 4t' to U is small we expand $E_{\rm vb}$ in powers of $\frac{U}{4t'}$ and get $E_{\rm vb} = -2t' + \frac{U}{2} - \frac{U^2}{16 t'}$. First term is the kinetic energy of two electrons occupying a bonding state. Second term is the Hartree term, coming from onsite repulsion U. Third term is the important correlation energy gain, that arises through a supression of double occupancy in the exact solution of the Hubbard dimer. This bonding between next nearest neighbor Bi sites (along diagonal Bi-S-Si direction) avoids nearest neighbor repulsion V.

\begin{figure}
\includegraphics[width=0.25\textwidth]{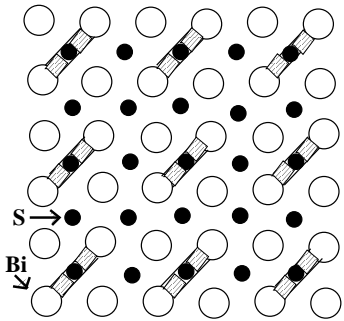}
\caption{A Wigner crystal of charge -2e valence bonds for \laofh, with commensurate doping x = 0.5. Inspite of valence bond localization its energy is lower than the free fermi state (figure 2), as this Wigner crystal completely avoids nearest neighbor repulsion V.} \label{fig4}
\end{figure}

Thus energy per Bi site of the frozen charge -2e Wigner crystal state is given by

\be
E_{\rm vbc} = \frac{1}{4} E_{\rm vb} \approx  - 0.3 eV
\ee

This energy is lower than the energy of the uniform shallow fermi sea state (equation 5), for our model Hamiltonian (equation 3). \textit{This completes our proof for the instability of uniform fermi sea towards charge -2e Cooper pair formation and Wigner crystallization.}

In proving an instability we have used a key nearest neighbor repulsion energy parameter V $\sim$ 1 eV. An accurate and microscopic evaluation of this parameter, relevant for low energy physics in \bis2 layers, as a function of doping x, in itself is a challenging manybody problem. Our parameter V, in some senses encodes all (non on-site) short distance unscreened electron-electron coulomb repulsions. 

Effective value of V for a given family member will depend on crystal and electronic structures, nature and dielectric properties of the reservoir layer etc. A large value of V we have assued is reasonable in an insulating state, such as a Wigner crystal we have discussed. Once the system gets metallized effective V will be modified. We believe that modifications at short distances will not be significant enough to alter our picture. 

Our main message is that in a low carrier density fermi sea unscreened short range repulsions can not be ignored, an insight that Wigner provided us back in 1934. In tight binding models with orbital degeneracies unscreened coulomb interaction seems to be even more important and has a potential to provide a rich physics, through self organization of generalized Wigner crystal, RVB states etc.

\section{Quantum Melting of Cooper Pair Wigner Crystal and Superconductivity}

Having demonstrated an instability of the uniform fermi sea to Wigner crystal formation, we study quantum fluctuations in the Cooper pair 
Wigner crystal. A given valence bond can undergo different quantum fluctuations, by electron hopping to available empty degenerate orbitals. 
Using the diagonal hopping t' a given X or Y-valence bond can i) expand its size by one diagonal unit in the diagonal (forward or backward)
 direction and ii) (pair) hop in the diagonal (forward or backward) direction by one unit. Our estimate of energy gain from these 
 processes is - $\frac{t'}{8}$. Further, an X valence bond can quantum tunnel to become a Y valence bond (a $\frac{\pi}{2}$ rotation) 
 using an off diagonal two body coulomb matrix element J$_{\rm p}$, taking advantage of non-zero overlap charges between the X and Y $\sigma$ bonding orbitals. Energy gain from this plaquette resonance (with d$_{xy}$ orbital symmetry, figure 5) 
per site is - $\frac{J_{\rm p}}{4}$. Since overlap charges are localized to the neigborhood of sulfur atoms, our estimate 
of   $J_{\rm p}$ is $\sim \frac{V}{4}$. 

In the above processes we also create virtual nearest neighbor occupancies. An estimate of coulomb energy cost per site from the above is $\frac{V}{8}$. Thus total energy gain from quantum fluctuations in the Cooper pair Wigner crystal state per Bi site is 
\be
- \frac{t'}{8} - \frac{J_{\rm p}}{4} + \frac{V}{8} \sim ~0.06~eV.
\ee
According to this estimate, a stable phase for the commensurate doping x = 0.5 is a Wigner crystal of plaqauete resonance bonds, or a Wigner crystal of charge -2e Cooper pairs,  shown in figure 5. It is also interesting to note the sign difference between the X and Y valence bonds, arising from a positive J$_p$ (pair hlpping between X and Y chains) indicates a d$_{xy}$ type of order parameter symmetry for superconductivity in this scenario.

\begin{figure}
\includegraphics[width=0.3\textwidth]{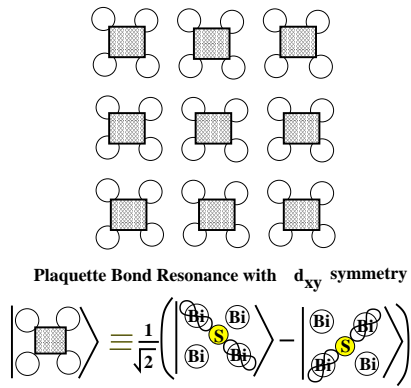}
\caption{Stating from state shown in figure 4, we allow for Bi-S-Bi diagonal valence bond to undergo $\frac{\pi}{2}$ rotational pair tunneling within the plaquettes to form an ordered plaquette bond resonance with d$_{xy}$ symmetry.
This Wigner crystalline state of charge -2e plaquette bond resonance gains a little more energy than charge -2e valence bond 
Wigner crystal state (figure 4).} \label{fig5}
\end{figure}

Additional quantum fluctuations might melt the charge -2e Cooper pair Wigner crystal even at the commensurate filling 
x = 0.5.  In existing experimental results for Tc as a function of x there is no dip seen around x = 0.5. 
This could indicate that the Cooper pair Wigner crystal has quantum melted and become a superconductor, even at x = 0.5.

As we move away from x = 0.5 we loose commensurability and Wigner crystal will quantum melt. Cooper pairs will also delocalize and expand their sizes, to take advantage of single particle delocalization. However, we expect the nature of valence bond correlations to survive at short distances even after quantum melting. That is, quantum melted Cooper pair fluid is a strongly correlated charge -2e fluid that avaoids nearest neighbor coulomb repulsions.
This as an interacting charge -2e bose fluid, a Kosterlitz-Thouless superconductor. Scale of superconducting Tc is determined by the \textit{effective band width of the delocalized charge -2e valence bond}. An unscreened near neighbor coulomb repulsions, the two fold orbital degeneracy and geometry of chemical bonding in the BiS lattice has stabilized this phase. 

Inview of the positive matrix element for hopping of valence bond between X and Y chains, as mentioned earlier we expect d$_{xy}$ symmetry for the spin singlet Cooper pair order parameter. This is likely to be a \textit{gapful d$_{xy}$ state} as fermi surfaces are fragmented. as seen in ARPES experiments \cite{arpesHHWen1,arpesHDingSmallFS,arpesOrbOrder,arpesFSTopology,arpesJapanese}. That is, nodal lines pass between fermi surfaces.

We estimate the effective band width of the valence bond pair using perturbation theory argument. An elementary valence bond pair motion involves single electron hopping, twice. In the intermediate state valence bond is broken. Perturbation theory estimate for valnce bond pair hopping is $
t_{vb} \approx \frac{t'^2}{- \frac{1}{2}[U^2 - 16t'^2]^{\frac{1}{2}} + \frac{U}{2}} \sim 0.7~eV$.
This is indeed a large energy scale. However, short range coulomb repulsion in this dense fluid of valence bond will renormalize the hopping matrix element significantly downwards and make Cooper pairs heavy. Further, fusion and regroupling of valence bonds through collisons will make the simple Bose fluid picture less and less relevant as we move away from the doping x = 0.5. This brings in exponential supression factor, similar to BCS and RVB mean field theories \cite{BZA}, giving us the following expression for Tc:

\be
K_BT_c \sim t_{\rm vb}  e^{\frac{-1}{J_{\rm eff} \rho(\epsilon_F)}}
\ee

Here J$_{\rm eff}$ is a mean energy required to break spin singlet pair and creat two electrons at the fermi level, for a short time scale. The exponential factor indicates a BCS like condensation of spin singlet spinon pairs, similiar to the situation in RVB mean field theory for cuprates \cite{BZA}. $\rho (\epsilon_{\rm F})$ is the density of states at the reference uniform fermi sea. When $\frac{U}{t}$ is large J$_{\rm eff}$ is approximated by the superexchange J $\approx\frac{4t^2}{U}$. However, for the present case of intermediate $\frac{U}{t}$, energy difference between the singlet ground state and excited triplet state of a Hubbard dimer is a good measure of $J_{\rm eff} \approx \frac{1}{2}[U^2 + 16t^2]^{\frac{1}{2}} - \frac{U}{2} \sim 1~eV $

This rough estimate gives us a maximum scale of Tc in the range 20 to 100 K.

A good way to study ground state properties of superconductivity quantitatively is to use the following variational RVB wave function:
\be 
|PRB\rangle_{\rm SC} = P_{\rm p} (\sum_{ij} \phi_{ij} b^\dagger_{ij})^{\frac{N_e}{2}}|0\rangle.
\ee
The function $\phi_{ij}$ represents Cooper pair function for a singlet valence bond electron pair at sites i and j. In terms of relative coordinates, amplitude $\phi_{ij}$ is large within a plaquette for diagonal singlets. A coherent superposition of diagonal singlets, within a plaquette represents the \textit{plaquette valence bond resonance} with d$_{xy}$ symmetry. The projection P$_p$ prevents overlap or touching of resonating plaquettes. This takes care of the short range coulomb repulsion and inturn overlap of valence bonds.

We defer study of the above variational wave function to a future publication and just point out that the short range repulsions in the BiS lattice, orbital degeneracy and a favorable hopping path way has given us an opportunity to have stable charge -2e Cooper pairs and a means to achieve superconductivity. 

In the next section we will see another nearly degenerate supercondducting state that can be obtained by creating quantum fluctuations in a charge -2e Wigner crystal.

\section{Transformation of 2d Cooper Pair Wigner Crystal into 1d array of Doped Mott-Hubbard Chains}

The superconducting and normal state that we are going to construct is interesting in its own right and has some phenomenological support as we will see. To realize this variational state we allow electrons to delocalize along only X chains, using hopping t', as shown in figure 6. This results in decoupled 1d chains, whose effective Hamiltonian within this manifold of X-orbital occupancy are decoupled 1d Hubbard X-chains. We start with a variational state denoted as (... , \half, 0, \half, 0, ...) to indicate a collection of parallel chains in their ground state with a staggered occupancy of \half and 0. As nearest neighbor occupancies are absent in  this variational state, inspite of delocalization along the chains, nearest neighbor repulsion V term drops out in the energy estimate.

\begin{figure}
\includegraphics[width=0.3\textwidth]{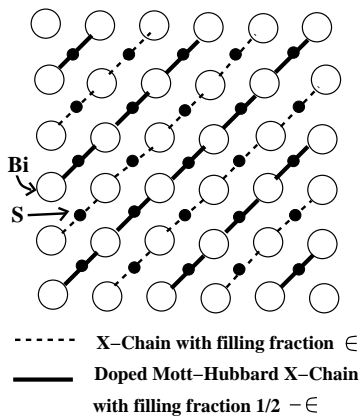}
\caption{We allow electrons in the X chains to delocalizes and form staggered array doped Mott-Hubbard chains
with occupancy $\frac{1}{2} -\epsilon$ and chains with a lower occupancy $\epsilon$. Energy of this state is lower than the uniform shallow fermi sea state of figure 2. We view this 1d array as a Wigner crystal of doped Mott-Hubbard chains.} \label{fig6}
\end{figure}

Energy of the above state at half filling is obtained using exact solutions of Lieb and Wu \cite{LiebWu}: 

\be
E_H(U) = -2t \int_{0}^{\infty} \frac{J_0(\omega) J_1(\omega)}{\omega(1 + \exp \frac{U}{2t}\omega)} d\omega~,
\ee
where J$_0(\omega)$ and J$_1(\omega)$ are Bessel functions.
This expression explicitly shows band narrowing effect or renormalization of hopping matrix element, as a function of U. For $\frac{U}{t'} \approx 2$, relevant for our problem, band narrowing is about 40 percent. Thus our interesting many body ground state has an energy per site:

\be 
E_c \approx - \frac{2t'}{\pi} \times \frac{3}{5} \approx 0.42 t' \approx - 0.3~eV
\ee.
 
Compared to the energy $\sim - 0.2$ eV  of free fermy gas with a full orbital and square lattice symmetry for the same value of x = 0.5, our correlated manybody state has a lower energy of $\sim -$ 0.1 eV per site.

We also find that we can reduce energy of the above state further (figure 5), by a staggered chain occupancy 
$( ..., \frac{1}{2} -\epsilon, ~~\epsilon~~,  \frac{1}{2} -\epsilon, ~~\epsilon~~,  \frac{1}{2} -\epsilon, ... )$, where the variational parameter $\epsilon \leq  0.25$. This state is a self doped state, where Mott insulating chains have transferred a fraction  $\epsilon$ of electrons to empty chains. We will discuss optimal value of $\epsilon$ etc., in a future publication, except to point out that for $\epsilon$ = 0.25 we have series of weakly coupled quarter filled bands, where only rotational symmetry is broken and lattice translation symmetry is preserved.

The above intrinsically anisotropic reference variational state allows us to construction anomalous normal states and superconducting states by making use of interchain hopping that we have ignored so far.

Mott insulating subsystes that we mostly encounter, for example in CuO or LaMnO$_3$ are robust Mott insulators. Mott insulating 3d electrons are robust organizations through strong chemical bonding of various atoms and after removal of orbital degeneracies. They are supported by the underlying lattice and a set of atoms.

In our case of generalized Wigner crystals, orbital selection and orbital order self organizes and supports doped Mott insulating chains with reduced symmetries. Doped Mott-Hubbard chain arrays breaks rotational and translational symmetry of the square lattice. There are 2$\times$2 = 4 possible chain orders: they are obtained by globally i) replacing x chains by y chains and ii) interchanging empty band chains and half filled band chains.

This leads to 4 different types of domain formation and accompanying phase boundaries, which are topological defects. Under suitable conditions these domains could order. Domain formation can be encouraged by disorder, for example arising from positional disorder of F atoms in \laofh. 

Energy gain in our low carrier density system by formation of doped Mott insulating state is small $\sim$ 0.1 eV. The self organized Mott chains can also disappear as a function of temperature, for entropic reasons. This will be a thermal melting of a generalized Wigner crystalline state.

\section{Interchain Pair Tunneling and Anisotropic 2D RVB Superconductivity}

We will sketch the mechanism of superconductivity in the weakly coupled doped Mott insulating chain scenario and make an estimate of superconducting Tc. Mott insulating chains develop a strong local spin singlet pairing correlations, as we move away from half filling. This is known in theory, from Lieb-Wu solution and numerical studies. Using the RVB theory for cuprates \cite{PWAScience87,BZA,BZAIran}, we suggest that neutral spin singlet pairs in the Mott-Hubbard chains are pre-existing Cooper pairs. In view of a finite Mott-Hubbard gap, singlet spinon pairs present in the ground state are not capable of transporting charges and remain \textit{neutral}. Doping makes room for a fraction of electron singlets to coherently delocalize and transport charge -2e. Fraction of charges that are capable of transporting charge -2e, the superfluid fraction, is proportional to the variational parameter $\epsilon$ (amount of self doping) or doping, $\delta \equiv$ 0.5 - x.

A doped Mott-Hubbard chain can not support true long range superconducting order, due to divergent 1d fluctuatuions. The interchain hopping can help in establishing a 2d Kosterlitz Thouless state. This is closely related to the interlayer pair tunneling  \cite{WHA} and interchain pair tunneling \cite{Kivelson} mechanism of superconductivity, which we briefly describe below.

A doped Mott-Hubbard chain, at low energies has spin-charge decoupling. It is described by Luttinger liquid theory. Coherent electron quasi particles are absent at the fermi surface. Consequently interchain hopping matrix elements get renormalized to zero at low energies \cite{StrongClarkeAnderson}.  It means that two conducting chains are effectively insulated and single electrons are confined within the chains. This is a many body effect, that was discussed in detail early in the context of interlayer pair tunneling mechanism in cuprates \cite{WHA,Kivelson,StrongClarkeAnderson}. However, if we consider a second order process, involving tunneling of pair of electrons, there can be coherent transport of a spin singlet charge 2e pair, using strong pairing correlations present in the individual chains. 

This coherent transport of charge 2e spin singlet pairs from chain to chain acts like a local Josephson tunneling. What is important is that energy gain from Josephson tunneling can i) increase local superconducting correlation and ii) establish Kosterlitz-Thouless superconductivity in 2 dimensions. In establishing a 2d superconductivity RVB superconducting correlations in the doped Mott chains play a fundamental role.

Expression for Tc for 2d superconductivity in interchain pair tunneling tunneling mechanism involves the pair tunneling matrix element $\frac{t''^2}{J_{\rm eff}}$:

\be
K_BT_c \sim \frac{t''^2}{J_{\rm eff}} e^{\frac{-1}{J_{\rm eff} \rho(\epsilon_F)}}
\ee

Here J$_{\rm eff}$ is a mean energy required to break spin singlet pair and creat two electrons at the fermi level, for a short time scale, so that they can pair tunnel via a second order process involving interchain hopping matrix element t'' etc. Meaning of the BCS like dependence is similar to the one we discussed in section V (following equation 9), for the isotropic 2d superconductivity. Substituting suitable numbers we get an estimate for maximum scale of  Tc  in the range $\sim$ 2 to 20 K.  This scale of Tc is much lower than the isotropic 2d state that we discussed in section V.

A well known place, where interchain pair tunneling seems to be at work, is in quasi 1d organic conductors \cite{Organic1D}.  Scale of interchain tunneling is typically 10 to 20 times lower. Interchain pair tunnelling has been also suggested \cite{Kivelson} to be a mechanism for pairing, when weakly coupled charge stripes are present in underdoped cuprates.

\section{Tunnling Asymmetry, Optical Conductivity, Bose Metal and Vortex Liquid}

Asymmetry in the tunneling conductance feature has been suggesed to be an important signal of an underlying Mott character in doped Mott insulating system \cite{TunnAsymOngAnderson}. Physically this asymmetry arises from differences in many electron reorganization, while adding or removing an electron in an STM experiment for example. The experimental work reported in reference \cite{stmStripe}, clearly shows a marked tunneling asymetry. In our theory, in Mott-Hubbard chain scenario as well as valence bond Wigner crystal scenario, we have strongly correlated Mott chain or Hubbard dimer which could produce the required asymmetry \cite{GBToBePublished} in tunneling conductance.

It will be nice to repeat the tunneling measurement for various doping level and study the asymmetry. Our expectation is that the asymmetry will be stronger at lower doping.

A very neat upper Hubbard band like feature is seen in optical conductivity around the energy 1.5 eV reported in \cite{Optics}. We suggest that this is a signal for the presence of Bi-S-Bi singlet pairs in the ground state. Energy required to break a valence bond pair in our theory (equation 6) is $E_{vb} \approx \sim$ 1.2 eV. Inclusion of residual coulomb interactions within our scenario could account for the peak seen at an energy of 1.5 eV in optical conductivity. More experiments will be welcome.

In the pseudo gap phase of cuprates pairing correlations or pre-existing RVB singlet pairs survive well above Tc.
In such a situation electrical transport, their magnetic field dependence and properties such as Nernst effect are dominated by Cooper pairs present at the fermi level and population of thermal vortices in the 2d CuO$_2$ planes. Anderson calls this phase as a vortex liquid \cite{vortexMetalPWAOng}. In another context, namely superconductor to insulator transition in 2d, for example induced by changing disorder, an intermediate metallic state domonated by Cooper pair like Bose correlations has been suggested. This is called Bose metal \cite{BoseMetal} by Das and Doniach. In our opinion there is a close connection between Bose metal and vortex liquid \cite{PrivateCommPWA}.

The theory we have developed in this paper for \bis2 offers a new play ground to achieve the physics of Bose metal and vortex liquid. In our model, pairing correlation arising from avoidance of next nearest neighbor repulsions, favors strong valence bond correlations with a large energy scale. This bosonic charge -2e correlation should survive in the normal state. Our suggestion is that the normal state of \bis2 family of superconductors should be studied carefully from the view of Bose metal and anomalous Nernst effect. 

In general the pseudogap features seen so for in various experiments should be investigated further to get deeper insight into the fascinating \bis2 family.

\section{Discussion and Some Predictions}

What is novel in our work is the organization of a dilute fermi sea in a band insulator, into
states containing certain diagonal valence bonds (charge -2e Cooper pairs). These states range from Cooper pair Wigner crystal to weakly coupled array of 1d Mott-Hubbard chains. The family of states we have suggested could support isotropic as well as anisotropic 2d Kosterlitz-Thouless superconductivity. It is likely both occur in different members of the \bis2 family for different range of parameters.

Self organization we have suggested, interestingly, makes onsite coulomb repulsion important, while trying to avoid near neighbor coulomb repulsions. It is a delicate interplay between onsite interaction and residual long range coulomb interactions, that leads to doped Mott physics and superconductivity.

In the Cooper pair Wigner crystal phase we have proposed, a disorder free \laofh is likely to be an insulator. In an ideal situation deviation away from x = 0.5 will act like a doping parameter. Disorder could quantum melt and create superconductivity even for x = 0.5. It will be interesting to see if the superconducting dome, currently observed around x = 0.5, develops a dip around x = 0.5 in superconducting Tc as a function of x, with decrease in disorder. 

Diagonal valence bond singlets Bi-S-Bi play a fundamental role in our theory. Topological defects in the generalized Wigner crystal may contain a finite density of unpaired spins (spinons) through broken diagonal valence bonds. This may lead to low temperature Curie contribution to spin susceptibility.

In the present paper we have only focussed on certain simple energy minimizing generalized Wigner crystal states. A closer look suggests that a variety of ordering of diagonal singlets, with closeby energies are possible.  One also expects glass like organization of the diagonal singlets for some range of parameters, leading to interesting electron pair glass or valence bond glass phases. As we indicated earlier, in the Mott-Hubbard chain scenario p-wave instability nematic metallic state etc. are interesting possibilities. 

Self organized states we have presented in this paper could be easily affected by disorder and temperature. So we expect interesting phase transitions as a function of temperature and disorder. As there is a spontaneous translation and orientational symmetry breaking associated with the self organization interesting topological defects, induced charges are possible. To test our theory it will be interesting to look for signals for enhanced antiferromagnetic (spin singlet) correlations in neutron and NMR experiments.  

Electronic coupling between two Bi-S layers in the \bis2 bilayers \cite{KurokiBilayer}, in our theory, could play important role in stabilizing different generalized Wigner crystals and their quantum melting properties. This needs to be investigated.

Instability towards the two phases we have talked about can also be justified using a weak coupling approach and a proper use of Umklapp electron pair scattering. They help build Mott localized spin singlet pairs \cite{GBToBePublished}. 

Our self organized Wigner crystal that supports doped Mott Hubbard chains or sheets is  different from the stripes one sees in underdoped cuprates and other Mott insulator based systems \cite{KivelsonStripe}. We organize strongly correlated systems starting from a fermi sea in a doped band insulator. Whereas stripe states in cuprates are reorganizations of an underlying Mott insulator after doping.

Electron phonon interaction will play very important role for small value of x. It is likely to stabilize isolated or islands, or even ordered arrays of Bi-S-Bi diagonal valence bonds. In the superconducting state, where valence bonds are delocalized, it is likely to play less important role. 

We hope to presemt, in a future publication,  a general theory of RVB states in doped band insulators from Coulomb forces and apply to other systems \cite{HfNCl,WO3,Diamond,Bi2Se3,STOLAO,gateSTO,gateMoS2}, HfNCl, WO$_3$, diamond, Bi$_2$Se$_3$ families, STO/LAO interface, gate doped SrTiO$_3$, MoS$_2$ etc., alluded to  at the beginning. 

\section{Summary}

In this article we discussed the puzzling phenomenology of the \bis2 layered family of superconductors and have suggest that they point towards a mechanism of superconductivity that is based on electron-electron repulsion.  Our main point is that utilizing the two fold orbital degeneracy at the fermi level and unscreened nearest neighbor coulomb repulsion, doped electrons get self organized into arrays of doped Mott chains or a quantum liquid or isotropic 2d RVB staates. We have called them Wigner crystal supported RVB states. Superconductivity, within our theory, can be either isotropic 2d or anisotropic 2d Kosterlitz-Thouless state.
In the isotropic (melted Wigner crystal) scenario order parameter symmetry is a gapful d$_{xy}$ symmetry; in the anisotropic (doped Mott Hubbard chain array) scenario we have a gapful extended-S symmetry.

We made an accurate estimate of energy of the vatiational state, using known exact results from Lieb and Wu for 1d repulsive Hubbard model. Rest of the conclusions, including a possible second phase containing plaquette bond resonances logically follow from this starting point. Our work at the moment is semi quantitative and suggestive.  It needs to be addressed more quantitatively. 

The novel scenario and a theory we have presented, that emphasizes \textit{importance of unscreened coulomb interactions in doped band insulators}, is encouraging from the point of reaching higher superconducting Tc's in this s-p, non transition metal based system. We hope it will raise debates, initiate further work and be of assistance to experimentalists to engineer synthesize of new members of the family, having reduced competing orders and higher superconducting Tc.

\textbf{Acknowledgement}. It is a pleasure to acknowledge V.P.S Awana, whose enthusiasm and continuing contrubtion to new superconductors encouraged me to think about the fascinating \bis2 layer family. I also thank S. Patnaik and S. Arumugam for discussion of their work. I am grateful to Science and Engineering Research Board (SERB, India) for a Fellowship. This work, partly performed at the Perimeter Institute for Theoretical  Physics, Waterloo, Canada is supported by the Government of Canada through Industry Canada and by the Province of Ontario through the Ministry of Research and Innovation.


\begin{thebibliography}{99}

\bibitem{BednorzMuller}J.G. Bednorz and A. Muller, Z. Phys. B: Condens. Matter {\bf 64} 188 (1986)

\bibitem{HfNCl} S. Yamanaka, K.I.Hotehama, H.Kawaji, Nature \textbf{392} 580 (1998)

\bibitem{WO3}S. Reich and Y.Tsabba, Eur. Phys. J. \textbf{B9} 1 (1999)

\bibitem{Diamond}E.A. Ekimov et al., Nature \textbf{428} 542 (2004);
G. Baskaran, J. Superconductivity and Novel Magnetism, \textbf{21} 45 (2008)

\bibitem{Bi2Se3} Y. S. Hor et al., Phys. Rev. Lett. \textbf{104} 057001 (2010)

\bibitem{STOLAO} Ariando et al., Nat. Comm. \textbf{2} 188 (2011)

\bibitem{gateSTO} K. Ueno et al., Nat. Mat. \textbf{7} 855 (2008) 

\bibitem{gateMoS2} K. Taniguchi et al., Appl. Phys. Lett. \textbf{101} 042603 (2012)

\bibitem{Mizuguchi2012} Y. Mizuguchi et al., J. Phys. Soc. Jpn. \textbf{81} 114725 (2012); Y. Mizuguchi et al., Phys. Rev. {\bf B 86} 220510 (2012)

\bibitem{Awana2012} K. Singh et al., J. Am. Chem. Soc. {\bf 134} 16504 (2012)

\bibitem{MizuguchiReview2014} Y. Mizuguchi, J. Phys. Chem. Solids,\textbf{ 84} 34 (2015)

\bibitem{BrianMappleReview} D. Yazici et al., Physica \textbf{C 514} 218 (2015)

\bibitem{Kuroki2012} H. Usui, K. Suzuki and K. Kuroko, Phys. Rev. {\bf B 86} 220501(R) (2012)

\bibitem{WignerCrystal} E. Wigner, Phys. Rev. \textbf{46} 1002 (1934)

\bibitem{LiebWu} E. H. Lieb and F. Y. Wu, Phys. Rev. Lett. {\bf 20} 1445 (1968)

\bibitem{PWAScience87} P.W. Anderson, Science, \textbf{235} 1196 (1987)

\bibitem{BZA} G. Baskaran, Z. Zou, P.W. Anderson, Solid St.  Com., \textbf{63} 973 (1987);
G. Baskaran, P.W. Anderson, Phys. Rev., \textbf{B 37} 580 (1988); P.W. Anderson et al.,  Phys. Rev. Lett., \textbf{58} 2790 (2007)

\bibitem{BZAIran} G. Baskaran, Iranian J. of Phys. Res., \textbf{6} 163 (2006)

\bibitem{WHA} J.M. Wheatley, T. Hsu and P.W. Anderson, Nature \textbf{333} 121 (1988)

\bibitem{Kivelson}V. J. Emery, S. A. Kivelson, and O. Zachar, 
Phys. Rev. \textbf{B 56} 6120 (1997)

\bibitem{StrongClarkeAnderson} David G. Clarke, S. P. Strong, and P. W. Anderson
Phys. Rev. Lett. \textbf{72} 3218 (1994) \

\bibitem{muSR1}G. Lamura et al., Phys. Rev. \textbf{B 88} 180509(R) (2013)

\bibitem{muSR2} P.K. Biswas et al., Physical Review \textbf{B 88} 224515 (2013)

\bibitem{SWavePatnaik}  Shruti, P Srivastava, and S Patnaik J. Phys.: Condens. Matter\textbf{ 25} 312202 (2013)

\bibitem{SWave2016} T. Yamashita et al.,  arXiv:1601.03502

\bibitem{BoseMetal}D. Das, D. and S. Doniach, Phys. Rev. \textbf{B 64} 134511 (2001);
P. Phillips, and D. Dalidovich, Science \textbf{302} 243 (2003)

\bibitem{vortexMetalPWAOng} P.W. Anderson, Nature Physics \textbf{3} 160 (2007)

\bibitem{PrivateCommPWA} G. Baskaran and P.W. Anderson (unpublished)


\bibitem{BCSPhonon1} I. R. Shein and A. L. Ivanovskii, JETP Letters, \textbf{96} 769 (2012)

\bibitem{BCSPhonon2} T. Yildirim, Phys. Rev. \textbf{B 87} 020506(R) (2013)

\bibitem{BCSPhonon3} X. Wan et al., Phys. Rev. \textbf{B 87} 115124 (2013)

\bibitem{BCSPhonon4} Y. Feng et al., J. Appl. Phys. \textbf{115} 233901 (2014)

\bibitem{BCSPhonon5} A. Athauda et al., arXiv:1601.07517
	
\bibitem{Elbio} G. B. Martins, A. Moreo, and E. Dagotto,  Phys. Rev. {\bf B 87} 081102(R) (2013)

\bibitem{ChineseTheory1}Y. Yang et al., Physical Review {\bf B 88}, 094519 (2013); C-L.Dai et al., Phys.Rev. {\bf B 91} 024512 (2015)

\bibitem{ChineseTheory2} X. Wu et al., Eur. Phys. Lett., {\bf 108} 27006 (2014)

\bibitem{ChineseTheory3} Yi Lang et al., Frontiers of Physics, {\bf 9} 194 (2014)

\bibitem{ChineseTheory4} Y. Gao, arXiv:1304.2102

\bibitem{SpinOrbit} Q. Liu et al., Phys. Rev. \textbf{B 91} 235204 (2015) 

\bibitem{KurokiBilayer} M. Ochi, R. Akashi and K. Kuroki, arXiv:1512.03884

\bibitem{chainBrazil} M.A. Griffith et al., arXiv: 1508.04820

\bibitem{arpesHHWen1} Z. R. Ye et al., Phys. Rev. {\bf B 90} 045116 (2014)

\bibitem{arpesHDingSmallFS} L. K. Zeng et al., Phys. Rev. \textbf{B 90} 054512 (2014)

\bibitem{arpesFSTopology} K. Terashima et al., Phys. Rev. {\bf B 90} 220512 (2014)

\bibitem{arpesOrbOrder} T. Sugimoto et al., Phys. Rev. \textbf{B 92} 041113 (2015)

\bibitem{arpesJapanese} N. L. Saini et al., Phys. Rev. {\bf B 90} 214517 (2014)

\bibitem{SoftXRay} S. Nagira et al., J. Phys. Soc. Japan \textbf{83} 033703 (2014)

\bibitem{stmStripe} T. Machida et al., J. Phys. Soc. Jpn. \textbf{83} 113701 (2014)

\bibitem{stmLargeGapHHWen} J. Liu et al., Euro. Phys. Lett., \textbf{106 } 67002 (2014) 

\bibitem{stmGapSpreadHHWen} S. Li et al., Sci. China-Phys. Mech. Astron \textbf{56 }2019 (2013)

\bibitem{TunnAsymOngAnderson} P.W. Anderson and N.P. Ong, J. Phys. Chem. Solids \textbf{67} 1 (2006) 

\bibitem{CDW250K} H.F. Zhai et al., Phys. Rev. \textbf{B 90} 064518 (2014)

\bibitem{neutron1} J. Lee et al., Phys. Rev. \textbf{B 87} 205134 (2013)

\bibitem{neutron2} A. Athauda et al., Phys. Rev. \textbf{B 91} 144112 (2015)

\bibitem{competingOrder}M. Fujita et al., Phys. Rev. Lett. \textbf{88} 167008 (2002)

\bibitem{GBCompetingOrder} G. Baskaran, Mod. Phys. Lett. \textbf{B 14} 377 (2000)

\bibitem{Doping1} H. Sakai et al., J. Phys. Soc. Jpn. \textbf{83} 014709 (2014)

\bibitem{Doping2} Y. Li et al., Supercond. Sci. Technol. \textbf{27} 035009 (2014) 

\bibitem{pressureMaple} Y. Fang et al.,  Phys. Rev.\textbf{ B 92} 094507 (2015)

\bibitem{pressureAwanaReview} R Jha and VPS Awana, arXiv:1512.08595

\bibitem{pressureArumugam} G. Kalai Selvan et al., Phys. Status Solidi RRL\textbf{ 7 }510 (2013)

\bibitem{pressureHHWen} J. Liu et al., Phys. Rev. \textbf{B 90} 094507 (2014)

\bibitem{uniaxialPressure} M. Fujioka et al.,	Eur. Phys. Lett. \textbf{108} 47007 (2014) 

\bibitem{Optics} X. B. Wang et al., Phys. Rev. \textbf{B 90} 054507 (2014)

\bibitem{Organic1D}C. Bourbonnais D. J´erome, in The Physics of Organic Superconductors and Conductors, Ed. A Lebed (Heidelberg: Springer 2008) p 357

\bibitem{GBToBePublished} G. Baskaran (to be published)

\bibitem{KivelsonStripe}E. Fradkin, S.A. Kivelson, and J.M. Tranquada, Rev. Mod. Phys., \textbf{87} 457 (2015).

\end{thebibliography}
\end{document}